\documentclass[11pt,letterpaper]{article}
\usepackage{graphicx}
\usepackage{tikz}
\usepackage{color}
\usepackage{ulem}
\usepackage{xcolor}
\usetikzlibrary{quotes,angles}
\usepackage{cite}

\usepackage[T1]{fontenc}
\usepackage[utf8]{inputenc}
\usepackage{authblk}

\title{Building Confidence in the Dirac $\delta$-function}

\author{Asim Gangopadhyaya\thanks{agangop@luc.edu}~}
\author{Constantin Rasinariu\thanks{crasinariu@colum.edu}}
\affil{Department of Physics, Loyola University Chicago}
\date{}

\begin{document}
\maketitle

\begin{abstract}
	In this note we present an example from undergraduate quantum mechanics designed to highlight the versatility of the Dirac $\delta$-function. Namely, we compute the expectation value of the Hamiltonian of a free-particle in a state described by a triangular wave function $\psi(x)$. Since the first derivative of $\psi(x)$ is piecewise constant, and because this Hamiltonian is proportional to the second order spatial derivative, students often end up finding the expectation value to be zero --an unphysical answer.  This problem provides a pedagogical application of the Dirac $\delta$-function. By arriving at the same result via alternate pathways, this exercise reinforces students' confidence in the Dirac $\delta$-function and highlights its efficiency and elegance.
\end{abstract}

\section*{Introduction}
Physics majors at most liberal arts colleges in United States generally see the Dirac $\delta$-function during their sophomore or junior year.  Its introduction is many times  associated with a disclaimer that the name  $\delta$-function is a misnomer because it is not a function; it should be seen as a distribution \cite{Aguirregabiria-2002,Boykin-2003}. However, as all practitioners know the essential role $\delta$-function plays in physics, it is important for the students to not only know its definition  \cite{Arfken-2001,Zhang-1989,Muller-1994,Ugincius-1972} and learn how it use it \cite{Atkinson-1975,Shankar-1994,Griffiths-2005}, but also to understand its subtlety \cite{Griffiths-1999}, and verify the results of using the $\delta$-function by alternate means, and thus incrementally build confidence in the robustness of the procedure.

Here, we present an example from undergraduate quantum mechanics \cite{Griffiths-2005} in which students are asked to determine the expectation value of a free particle Hamiltonian $\hat{H}$ in a particular state $\psi(x)$ by several methods, including one that employs the Dirac $\delta$-function.  

\section*{Description of the Problem} 
\begin{figure} [htb]
	\centering
	\includegraphics[width=0.55\linewidth]{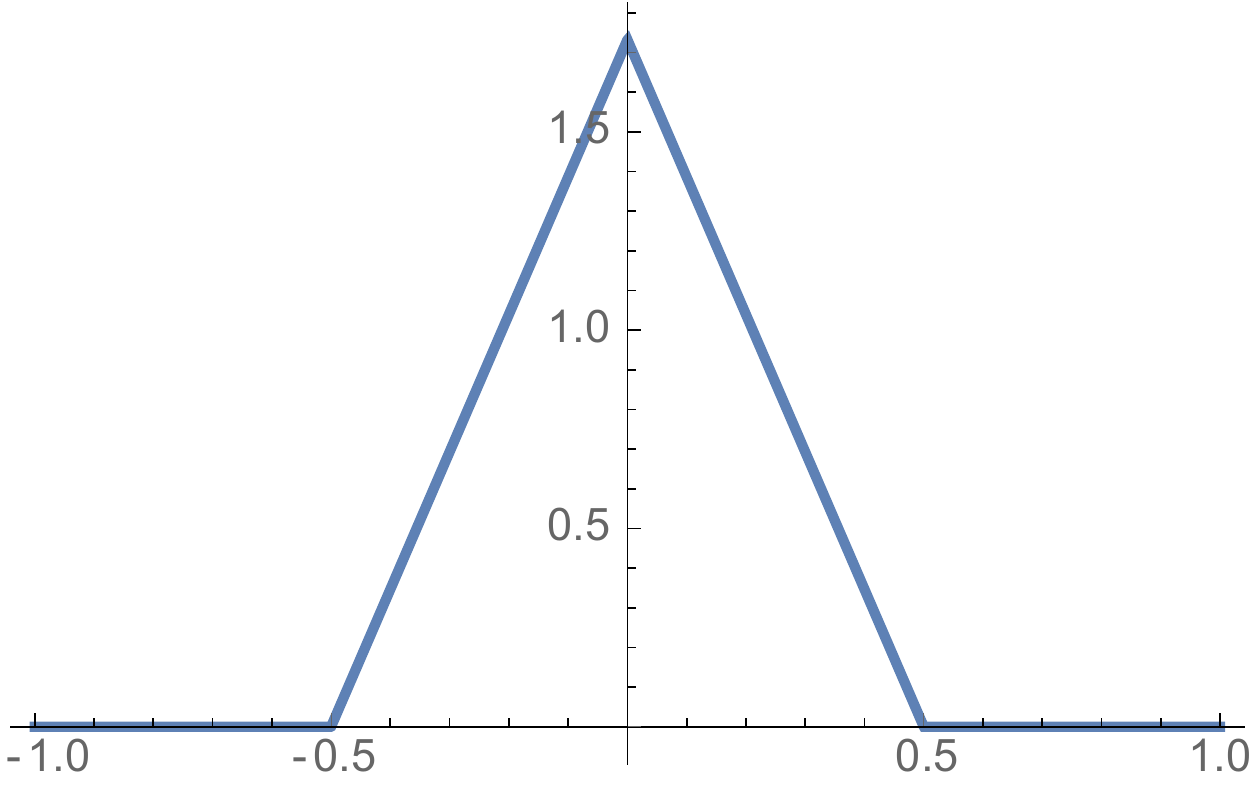}
	\caption{An isosceles triangular wave function of height $\sqrt{\frac3{a}}$ and extending from $-a/2$ to $+a/2$ for $a=1$.}
	\label{fig:fx}
\end{figure}

Consider a free particle, which at $t=0$, is in a state described by the following normalized triangular wave function
\begin{equation}
\psi(x) = \left\lbrace 
\begin{array}{ll}
{2\sqrt{\frac3{a^3}}\,(x+\frac a2)}		& \quad\quad -\frac a2 < x <0\\&\\
{2\sqrt{\frac3{a^3}}\,(\frac a2-x)}	& \quad\quad\quad \! 0\leq x <\frac a2\\&\\
0 & \quad\quad\quad  |x|\geq\frac a2
\end{array}
\right. \label{eq:fx}
\end{equation}
The isosceles triangle with its base from $-a/2$ to $+a/2$, illustrated in Fig. (\ref{fig:fx}), is a square-integrable function, and hence is a valid wave function for any potential defined over a segment of the real-axis (or the entire real-axis) containing $[-a/2,+a/2]$. 
It is interesting to note that such a function is also the groundstate for an infinite square well potential with an attractive $\delta$-potential in the middle \cite{Ahmed-2014}, and for such a system its shape does not evolve with time.

The Hamiltonian for the system is $\hat{H}=-\frac{\hbar^2}{2m}\frac{\partial^2}{\partial x^2}$. Our objective is to determine the expectation value of the Hamiltonian in this state.

In our experience, when presented with this problem,  many times students rush to the naive conclusion that $\langle H \rangle = -\frac{\hbar^2}{2m} \int \psi^*(x) \frac{\partial^2}{\partial x^2}\, \psi(x)\, {\rm d}x = 0$, because the first derivative of the wave function is piecewise constant. They would soon realize that, because of the discontinuities of $\psi'(x)$, the second derivative of the wave function must introduce several Dirac $\delta$-functions. This allows them to calculate the integral in a straightforward way. However, we think that students' understanding improves when alternate, equivalent solution pathways are presented. Therefore, in this paper we will use three different methods to compute the expectation value of $\langle H \rangle $. First, we consider that the free particle is inside an infinite square well of width $a$. Thus, $\psi(x)$ can be expressed in terms of the eigenfunctions $\phi_n(x)$ of the square well Hamiltonian. This is equivalent to the Fourier series representation of $\psi(x)$. The second method makes use of the Fourier transform of $\psi(x)$ which, up to a constant $\hbar$, gives the momentum wave function representation $\phi(p)$. The third method uses the Hermiticity of the momentum operator \cite{Ahmed-2014a} allowing us to cast $\langle \psi |\hat p^2|\psi\rangle$ as $\langle \hat p \psi |\hat p \psi\rangle$. This turns out to be the easiest way to compute the integral. Finally, the fourth method employs the Dirac $\delta$-function mentioned above, and is by far the most efficient way of getting the result. We believe that by solving the same problem via  several equivalent ways,  students become more comfortable in using the Dirac $\delta$-function and, equally important, learn to appreciate  the elegance and effectiveness of such techniques.

\section*{Method 1: Fourier Series} 

In this section we assume that the particle is free within the domain $(-\frac a2, \frac a2)$ and is blocked from the  beyond by infinite barriers at $x=\pm \frac a2$. The eigenfunctions are $\phi_{n}(x)=\sqrt{\frac2a} \, \cos (n\pi x/a)$ for odd $n$'s and $\phi_{n}(x)=\sqrt{\frac2a} \, \sin (n\pi x/a)$ for even $n$'s, where $n=1,2,3,\cdots$.  As expected, they vanish  at $x=\pm \frac a2$.  Since $\phi_{n}(x)$ form a complete set, we can expand the triangular function $\psi(x)$ as the linear combination
$$
\psi(x) = \sum_{n=1}^\infty c_n \,\phi_{n}~.
$$
%From the symmetry of $\psi(x)$, we see that all $c_{2n}$=0. Hence,
Since $\psi(x)$ is an even function, only the cosine functions remain, with odd coefficients $c_{2n+1}$. Hence,
$$
\psi(x) = \sum_{n=0}^\infty c_{2n+1} \,\left( \sqrt{\frac2a} \, \cos \left[ \frac{(2n+1)\pi x}{a}\right]\right) ~.
$$
The coefficients $c_{2n+1}$ are given by
\begin{eqnarray*}
c_{2n+1} 	&=& \int_{-a/2}^{a/2} \,(\phi_{n})^* \psi(x) dx~\\
			&=& 2  \int_{-a/2}^{0} \,\left(2 {\sqrt{\frac3{a^3}}}\,\cdot\, \left( x+\frac a2\right) \right) \, \left( \sqrt{\frac2a} \, \cos \left[ \frac{(2n+1)\pi x}{a}\right]\right)   		dx~\\
			&=& \frac{4\sqrt{6}}{\pi^2 \,(2n+1)^2} ~.
\end{eqnarray*}
The expectation value of the Hamiltonian is
\begin{eqnarray}
\left\langle \frac{\hat{p}^2}{2m}\right\rangle 	
			&=& -\frac{\hbar^2}{2m}\int_{-a/2}^{a/2} \,(\psi(x))^* ~~ \left( \frac{d^2\psi}{dx^2}\right)  dx~			
			\nonumber \\ 
			&=& \frac{\hbar^2}{2m}\sum_{k,n=0}^\infty c_{2n+1}\, c_{2k+1} 
			\left( \frac{(2k+1)\pi}{a}\right)^2  
			\nonumber \\ 
			&&\qquad \times
			\underbrace{\int_{-a/2}^{a/2} \,
			\sqrt{\frac2a}\cos\left( \frac{(2n+1)\pi\,x}{a}\right)  \cdot \sqrt{\frac2a}\cos\left( \frac{(2k+1)\pi\,x}{a}\right) dx}_{\delta_{n,k}} \nonumber \\ 
			&=& \frac{\hbar^2}{2m}\, \sum_{n=0}^\infty\left( \frac{4 \sqrt{6}}{\pi^2 \,(2n+1)^2} \right)^2
			\left( \frac{(2n+1)\pi}{a}\right)^2
			\nonumber \\ 			
			&=& \frac{96\hbar^2}{2m\pi^2\,a^2}\,\cdot\, \sum_{n=0}^\infty  \frac{1}{ \,(2n+1)^2} 
			~=~ \frac{96\hbar^2}{2m\pi^2\,a^2}~\cdot~\frac{\pi^2}8 
			~=~ \frac{6\hbar^2}{ma^2}~,
\end{eqnarray}
where we used Mathematica \cite{Mathematica-11.3} to compute the infinite sum
$$
\sum_{n=0}^\infty  \frac{1}{ \,(2n+1)^2}  = \frac{\pi^2}8 .
$$
Students can arrive at the same result using the formula 0.234.2 on page 7 of \cite{Gradshteyn-1979}.

\section*{Method 2: Fourier Transform}

\begin{figure}
	\centering
	\includegraphics[width=0.55\linewidth]{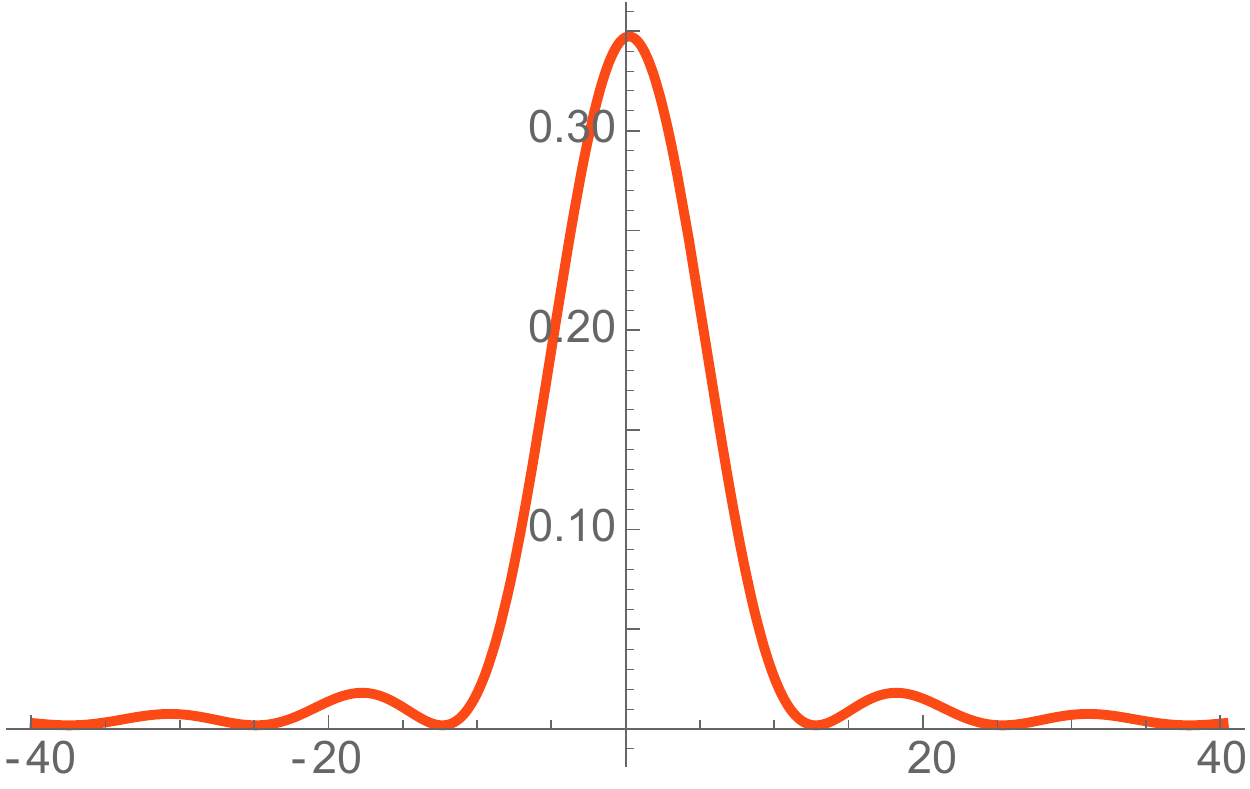}
	\caption{The momentum space wave function for $a=1$ and $\hbar=1$.}
	\label{fig:fp}
\end{figure}

In this section we will use the momentum wave function representation $\phi(p)$ which, up to a constant $\hbar$, is exactly the inverse Fourier transform of $\psi(x)$:
\begin{equation}
	\phi(p) = \frac{1}{\sqrt{2\pi\hbar}} \int_{-\infty}^{+\infty} \psi(x)
	e^{ -ipx/\hbar }\, dx~.
\end{equation}
For the triangular wave function given in Eq. (\ref{eq:fx}), we obtain
\begin{eqnarray}
\phi(p) &=& 
\frac{2}{\sqrt{2\pi\hbar}}\sqrt{\frac{3}{a^3}}\left( 
\int_{-a/2}^{0} (x+a/2) e^{ -ipx/\hbar } dx + \int_{0}^{a/2}(a/2-x)  e^{ -ipx/\hbar } dx 
\right) \nonumber \\
&=& \frac{1}{p^2}\sqrt{\frac{6\hbar^3}{\pi a^3}} 
\left(
2- e^{ -iap/2\hbar }- e^{ iap/2\hbar }
\right)
\nonumber 
\\
&=&4 \sqrt{\frac{6 {\hbar}^3}{\pi a^3}}
\times \frac{1}{p^2} \sin ^2\left(\frac{a p}{4 {\hbar}}\right)~. \label{eq:MSWF}
\end{eqnarray}
In Fig. (\ref{fig:fp}), we depict the momentum space wave function $\phi(p)$.
In this representation, the expectation value of the Hamiltonian is 
\begin{equation}
	\left\langle\frac{\widehat{p}^2}{2m} \right\rangle = \frac{1}{2m}\int_{-\infty}^{+\infty} p^2 \left|\phi(p)\right|^2\, dp~.
\end{equation}
Now, inserting $\phi(p)$ from Eq. (\ref{eq:MSWF}) into the above equation, we get
\begin{equation}
\left\langle  \frac{\widehat{p}\,^2}{2m}\right\rangle = 
\frac{96 \hbar^3}{2m \pi a^3} \int_{-\infty}^{+\infty}
\frac{1}{p^2} \sin ^4\left(\frac{a p}{4 {\hbar}}\right)\,dp = \frac{6\hbar^2}{ma^2}
\end{equation}
Again, we resorted to Mathematica \cite{Mathematica-11.3} to calculate the integral 
$$ \int_{-\infty}^{+\infty}
\frac{\sin ^4\left(x\right)}{x^2}\,{\rm d}x = \frac\pi 2~~. $$
Equivalently, one can use the formula 3.821.10 from page 446 of \cite{Gradshteyn-1979}.
\section*{Method 3: Hermiticity of the Momentum Operator}
Since the momentum operator $\hat p$ is Hermitian, we can write $\langle \psi |\hat p^2|\psi\rangle$ as $\langle \hat p \psi |\hat p \psi\rangle$. Thus, in coordinate representation, we have 
\begin{eqnarray}\left\langle  \frac{\widehat{p}\,^2}{2m}\right\rangle	&=&
\frac{1}{2m}\langle \hat p \psi |\hat p \psi\rangle	\nonumber\\
&=& \frac{1}{2m} \, \int_{-a/2}^{a/2} \,\left( -i \hbar\, \frac{d\psi}{dx}\right)^* ~ \left( -i \hbar\,\frac{d\psi}{dx}\right)  dx, \nonumber\\
&=& \frac{\hbar^2}{2m} \, \left( 2\sqrt{\frac3{a^3}}\right)^2 \int_{-a/2}^{a/2} dx
\nonumber\\
&=& \frac{6\hbar^2}{ma^2} 
\label{eq:H_by_Hermiticity}
\end{eqnarray}
This method, used in Ref. \cite{Ahmed-2014a}, turns out to be the easiest of the four methods considered here.

\section*{Method 4: Dirac $\delta$-function}
Since our objective is to see an application of the $\delta$-function, in this section we will derive the expectation value by directly applying $\hat{H}$ to the triangular wave function of Eq. (\ref{eq:fx}).
\begin{eqnarray}\left\langle  \frac{\widehat{p}\,^2}{2m}\right\rangle	
			&=& -\frac{\hbar^2}{2m} \, \int_{-a/2}^{a/2} \,(\psi(x))^* ~ \left( \frac{d^2\psi}{dx^2}\right)  dx~~.\label{eq:H_by_delta}
\end{eqnarray}	
To compute this integral, we first find $\frac{d\psi}{dx}$ and  $\frac{d^2\psi}{dx^2}$. The first derivative is given by
\begin{equation}
\frac{d \psi(x)}{dx} = \left\lbrace 
\begin{array}{ll}
2\sqrt{\frac3{a^3}}		& \quad\quad -\frac a2 < x <0\\&\\
-2\sqrt{\frac3{a^3} }	& \quad\quad\quad \! 0< x <\frac a2\\&\\
0 & \quad\quad\quad  |x|\ge\frac a2
\end{array}
\right. \label{eq:dfx}
\end{equation}

\begin{figure}[htb]
	\centering
	\includegraphics[width=\linewidth]{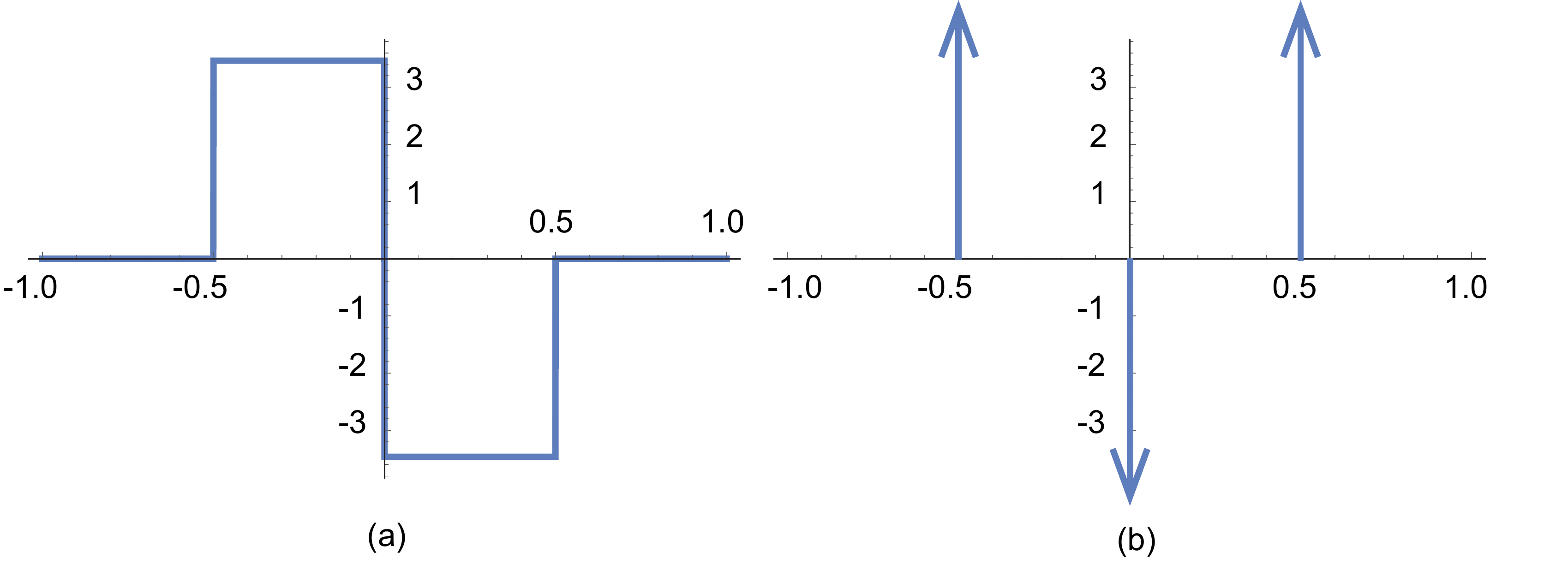}
	\caption{The first derivative (a) and second derivative (b) of $\psi(x)$ for $a=1$. The $\delta$-functions are represented by vertical arrows.}
	\label{fig:d}
\end{figure}

\noindent
Due to the discontinuities of $\frac{d\psi}{dx}$ at $x=-a/2, x=0$ and $x=a/2$, the second derivative will be a sum of three $\delta$-functions centered on these points:
\begin{equation}
\label{eq:delta}
\frac{d^2\psi}{dx^2} = 2\sqrt{\frac3{a^3}}~ \delta\left( x+\frac a2\right)
-4\sqrt{\frac3{a^3}}\,~ \delta(x)
+2\sqrt{\frac3{a^3}}~ \delta\left( x-\frac a2\right)~.
\end{equation}
Note that the coefficients in front of each $\delta$-function are equal to the differences in the derivatives of $\psi$ across each point of discontinuity. 

The first and second derivatives are depicted in Fig. (\ref{fig:d}) for $a=1$. Inserting Eq. (\ref{eq:delta}) into Eq. (\ref{eq:H_by_delta}), and taking into accont that $\psi\left( \pm\frac a2\right) = 0$, we get
\begin{eqnarray}\left\langle  \frac{\widehat{p}\,^2}{2m}\right\rangle			&\!\!\!\!\!\!\!\!=\!\!\!\!\!\!\!\!&  -\frac{\hbar^2}{2m} \, \int_{-a/2}^{a/2} \,\psi^*(x) 
			\left[2\sqrt{\frac3{a^3}}~ \delta\left( x+\frac a2\right)
			-4\sqrt{\frac3{a^3}}\,~ \delta(x)
			+2\sqrt{\frac3{a^3}}~ \delta\left( x-\frac a2\right)
			\right]  dx~ \nonumber \\ 	\nonumber \\ 		
			&\!\!\!\!\!\!\!\!=\!\!\!\!\!\!\!\!& -\frac{\hbar^2}{2m} \, \left[  \psi^*\left( -\frac a2\right) ~\left(2\sqrt{\frac3{a^3}}~ \right)  
			-(\psi(0))^* ~\left(-4\sqrt{\frac3{a^3}}~ \right) 
			 -\psi^*\left( \frac a2\right)\left(2\sqrt{\frac3{a^3}}~ \right) \right] 
			\nonumber \\ 	\nonumber \\ 		
			&\!\!\!\!\!\!\!\!=\!\!\!\!\!\!\!\!&-\frac{\hbar^2}{2m} \cdot  \left( 2 {\sqrt{\frac3{a^3}}}\,\cdot\, \frac a2
			\right)  \left(-4\,\sqrt{\frac3{a^3}}~\right) \nonumber\\ 		
			&\!\!\!\!\!\!\!\!=\!\!\!\!\!\!\!\!& \frac{6\hbar^2}{ma^2}~.
			 \end{eqnarray}
Note that this derivation is also straightforward and, unlike the first two methods, doesn't require any use of Mathematica.

%The equality of these methods also raises our confidence in the completeness of the $\phi_{n}(x)=\sqrt{\frac2a} \, \cos (n\pi x/a)$ in the space of even functions.
\section*{Conclusion}
In this note we computed the expectation value of a free-particle Hamiltonian in a state described by a triangular wave function $\psi(x)$. Since the first derivative of $\psi(x)$ is piecewise constant, and because the Hamiltonian is proportional to a second order spatial derivative, students often end up finding the expectation value to be zero --an unphysical answer.  This problem provides a natural setting for the introduction of Dirac $\delta$-function as the derivative of a step function. By computing the expectation value of the Hamiltonian via several different methods we attempt to bolster their confidence in $\delta$-function, which appears in many areas of physics. 

\section*{Acknowledgement} We would like to thank the reviewers for their valuable comments and express our gratitude to the editors for the opportunity to improve our manuscript.


\begin{thebibliography}{99}
%\bibitem{Gallitto-2011}
%
\bibitem{Aguirregabiria-2002}
Aguirregabiria J M, Hern\'andez A, and Rivas M 2002, $\delta$-function converging sequences,
Am. Jour. of Phys. {\bf 70}, 180; doi: 10.1119/1.1427087
%
\bibitem{Boykin-2003}
Boykin T B 2003, Derivatives of the Dirac delta function by explicit construction of sequences, Am. Jour. of Phys. {\bf 71}, 462; doi: 10.1119/1.1557302
%
\bibitem{Arfken-2001} Arfken G B and Weber H J 2001, \textit{Mathematical Methods for Physicists} 5th Ed. (Harcourt).
%
\bibitem{Zhang-1989} Zhang D, Ding Y and Ma T 1989, A note on the definition of delta functions, Am. Jour. of Phys. {\bf 57}, 281; doi: 10.1119/1.16058
%
\bibitem{Muller-1994}
Muller F A 1994, Definitions of Delta Distributions, Am. Jour. of Phys. {\bf 62}, 11; doi: 10.1119/1.17729
%
\bibitem{Ugincius-1972}
Ugin\v{c}ius P 1972, An Integral Representation for the Dirac Delta Function, Am. Jour. of Phys. {\bf 40}, 1690; doi: 10.1119/1.1987016
%
\bibitem{Atkinson-1975}
Atkinson D A, and Crater H W 1975, An exact treatment of the Dirac delta function potential in the Schrödinger equation, Am. Jour. of Phys. {\bf 43}, 301; doi: 10.1119/1.9857
%
\bibitem{Shankar-1994} Shankar R 1994, \textit{Principles of Quantum Mechanics} 2nd Ed. (Springer).
%
\bibitem{Griffiths-2005} Griffiths D J 2005, \textit{Introduction to Quantum Mechanics} 2nd Ed. (Pearson Prentice Hall).
%
\bibitem{Griffiths-1999}
Griffiths D, and Walborn S 1999, Dirac deltas and discontinuous functions,
Am. Jour. of Phys. {\bf 67}, 446; doi: 10.1119/1.19283
%
\bibitem{Ahmed-2014} Ahmed Z and Kesari S, The simplest model of the zero-curvature eigenstate, Eur. J. Phys. {\bf 35}, 2014; doi: doi:10.1088/0143-0807/35/1/018002
%
\bibitem{Ahmed-2014a} Ahmed Z and Yadav I, Position-momentum uncertainty products, Eur. J. Phys. {\bf 35}, 2014;  doi:10.1088/0143-0807/35/4/045015
%
\bibitem{Mathematica-11.3} Wolfram Research, Inc., Mathematica, Version 11.3, Champaign, IL (2018).
%
\bibitem{Gradshteyn-1979} Gradshteyn I S and Ryzhnik I M 1979, \textit{Tables of Integrals, Series, and Products}, Corrected and Enlarged Ed, (Academic Press)
	
\end{thebibliography}
\end{document}